\documentclass[12pt]{iopart}

\usepackage[english]{babel}
\usepackage[dvips]{graphicx}
\usepackage[dvips]{color}
\usepackage{amssymb}

\begin{document}

\newcommand{\mtc}[1]{\mathcal{#1}}
\newcommand{\mtr}[1]{\mathrm{#1}}
\newcommand{\mtb}[1]{\mathbf{#1}}
\newcommand{\nad}{n_{\rm{ad}}}
\newcommand{\nadI}{n_{\rm{ad1}}}
\newcommand{\nadII}{n_{\rm{ad2}}}
\newcommand{\niso}{n_{\rm{iso}}}
\newcommand{\ncor}{n_{\rm{cor}}}
\newcommand{\acor}{\alpha_{\rm{cor}}}

\newcommand{\abs}[1]{\vert #1 \vert}

\title{Hints of Isocurvature Perturbations in the Cosmic Microwave Background?}

\author{Reijo Keskitalo$^{1,2}$, Hannu Kurki-Suonio$^{2}$, Vesa Muhonen$^{1,2}$ and
  Jussi V\"{a}liviita$^{3}$}

\address{$^{1}$ Helsinki Institute of Physics, University of Helsinki, P.O.
  Box 64, FIN-00014 Helsinki, Finland}

\address{$^{2}$ Department of Physical Sciences, University of Helsinki, P.O.
  Box 64, FIN-00014 Helsinki, Finland}

\address{$^{3}$ Institute of Cosmology and Gravitation, University of
  Portsmouth, Portsmouth PO1 2EG, United Kingdom}

\begin{abstract}
  The improved data on the cosmic microwave background (CMB) anisotropy allow
  a better determination of the adiabaticity of the primordial perturbation.
  Interestingly, we find that recent CMB data seem to favor a contribution of
  a primordial isocurvature mode where the entropy perturbation is positively
  correlated with the primordial curvature perturbation and has a large
  spectral index ($\niso \sim 3$).  With 4 additional parameters we obtain a
  better fit to the CMB data by $\Delta \chi^2 = 9.7$ compared to an adiabatic
  model. For this best-fit model the nonadiabatic contribution to the CMB
  temperature variance is 4\%. According to a Markov Chain Monte Carlo
  analysis the nonadiabatic contribution is positive at more than 95\% C.L.
  The exact C.L. depends somewhat on the choice of priors, and we discuss the
  effect of different priors as well as additional cosmological data.
\end{abstract}



\maketitle

\section{Introduction}

A fundamental question in cosmology is the origin of the density perturbation,
from which the structure of the universe (galaxies, and galaxy clusters) has
grown. Today a popular scenario is inflation, where the perturbation
originates as a quantum fluctuation of the inflaton field.

Clues to the origin of structure can be sought in the nature of this
perturbation. Simple inflation models produce adiabatic perturbations, but
more complicated (multi-field) models may also produce nonadiabatic
perturbations.

Current observations are consistent with adiabatic primordial perturbations,
but sizable deviations from adiabaticity remain allowed. Here ``primordial''
refers to the early radiation-dominated epoch (e.g., at some time soon after
big bang nucleosynthesis), when all cosmological scales ($\gtrsim 1$~Mpc) are
well outside the horizon. Adiabatic perturbations remain adiabatic while
outside the horizon, but give rise to entropy perturbations as they enter the
horizon. Adiabatic perturbations are completely characterized by the
associated (comoving gauge) curvature perturbation $\mtc{R}$, whereas
nonadiabatic perturbations have entropy perturbations $ \mtc{S}_{ij} \equiv
\delta_i/(1+w_i) - \delta_j/(1+w_j)$ between different constituents $i$, $j$
to the energy density. Here $\delta$ is the dimensionless relative density
perturbation and $w \equiv p/\rho$ is the ratio of pressure to energy density.

A general perturbation can be divided into an adiabatic mode and a number of
isocurvature modes, which evolve independently. The adiabatic mode has
$\mtc{S}_{ij} = 0$ initially (i.e., outside the horizon in the
radiation-dominated epoch), whereas isocurvature modes have $\mtc{R} = 0$ and
$\mtc{S}_{ij} \neq 0$ initially.  There are four different types of
isocurvature perturbations \cite{Bucher:1999re} --- the cold dark matter
(CDM), baryon, neutrino density, and neutrino velocity isocurvature modes. For
simplicity, we consider here only the CDM mode, with an initial entropy
perturbation
\begin{equation}
  \mtc{S} \equiv \delta_{c} - \frac{3}{4}\delta_{\gamma} \,.
\end{equation}
The baryon mode is observationally very similar.

Even though purely isocurvature perturbations have been ruled out
\cite{Enqvist:2001fu, Enqvist:2000hp}, the data allow a subdominant
($\sim$10\% level \cite{Kurki-Suonio:2004mn}) isocurvature contribution.
Depending on how the primordial $\mtc{R}$ and $\mtc{S}$ perturbations were
generated, they may be correlated with each other \cite{Langlois:1999dw}.

Observationally the CDM isocurvature mode differs from the adiabatic mode in
the locations of the acoustic peaks in the CMB angular power spectrum
$C_{\ell}$. The presence of an isocurvature contribution, especially a
correlated one, appears in $C_{\ell}$ as a change in the ratio of peak
separation to the first peak position. Since the first peak is well fixed by
the present data, an isocurvature contribution would appear as a shift in the
position of the other peaks, i.e., as a reduction in peak separation.  CMB
measurements of increasing accuracy allow thus a tighter constraint on the
isocurvature contribution. For studies utilizing the Wilkinson Microwave
Anisotropy Probe (WMAP) 1-year data \cite{Bennett:2003bz}, see
\cite{Kurki-Suonio:2004mn, Valiviita:2003ty, Beltran:2005xd, Peiris:2003ff,
  Crotty:2003rz, Parkinson:2004yx, Moodley:2004nz, Ferrer:2004nv,
  Beltran:2004uv, MacTavish:2005yk}.

The WMAP 3-year data \cite{Spergel:2006hy} is an improvement in this respect
\cite{Bean:2006qz, Lewis:2006ma, Trotta:2006ww}. While the first peak was
already measured very accurately in the first year, the determination of the
second peak shape and location is improved with the 3-year data. Likewise the
new Boomerang data \cite{Jones:2005yb} begins to define the third peak. This
motivates a new study to update our earlier results
\cite{Kurki-Suonio:2004mn}.

The focus of this paper is on what \emph{the CMB data} say about the nature of
primordial perturbations. Thus we use CMB and large-scale structure (LSS) data
only, but address other cosmological data in the end. Inclusion of LSS data
was needed to break certain parameter degeneracies \cite{Tegmark:2006az} and
to constrain extreme values for spectral indices. Current CMB data do not
cover with good accuracy a sufficient range of scales to constrain well the
several independent spectral indices of our model. The Planck satellite will
eventually fix this situation.

\section{Model}

We consider a flat ($\Omega_0 = 1$) $\Lambda$CDM model with primordial
curvature and entropy perturbations, which may be correlated.  For details of
our model, see \cite{Kurki-Suonio:2004mn}.  We give below the main points.

We divide the primordial curvature perturbation into an uncorrelated and a
fully correlated part. We assume the power spectra of these perturbations and
correlations can be characterized by power laws, but allow different spectral
indices for the entropy and curvature perturbation. Thus the spectra can be
written as
\begin{equation}
  \label{eq:correlations}
  \eqalign{
    \mtc{P}_{\mtc{R}}(k) \equiv \mtc{C}_{\mtc{RR}}(k) =
    A^{2}_{r} \hat{k}^{\nadI-1} + A^{2}_{s} \hat{k}^{\nadII-1}, \\
    \mtc{P}_{\mtc{S}}(k) \equiv \mtc{C}_{\mtc{SS}}(k) = B^{2} \hat{k}^{\niso-1}, \\
    \mtc{C}_{\mtc{RS}}(k) = \mtc{C}_{\mtc{SR}}(k) = A_{s}B \hat{k}^{\ncor-1},}
\end{equation}
where $\hat{k} = k/k_{0}$ and $k_0 = 0.01 \mbox{Mpc}^{-1}$ is the pivot scale
at which the amplitudes are defined. The spectral index $\ncor$ is not an
independent parameter and is defined by $\ncor = (\nadII + \niso)/2$. Our
approach is phenomenological, looking for signatures of deviation from
adiabaticity in the observations, and is not tied to a particular model for
generating the primordial perturbations. The main assumption is that the
spectra can be approximated by constant spectral indices over the range of
distance scales probed by the data. For motivation and discussion of our
parametrization in the inflation context, see \cite{Kurki-Suonio:2004mn,
  Gordon:2000hv, Amendola:2001ni, Byrnes:2006fr}.

The total CMB spectrum $C_{\ell}$ can now be divided into four components: the
uncorrelated adiabatic part, the correlated adiabatic part, the isocurvature
part, and the correlation between the last two \cite{Kurki-Suonio:2004mn}:
\begin{equation}
  \label{eq:total-Cl}
  \eqalign{
    C_{\ell} &= A^{2}\bigr[ (1-\alpha)(1-\abs{\gamma})
             \hat{C}^{\mtr{ad1}}_{\ell}
             + (1-\alpha)\abs{\gamma} \hat{C}^{\mtr{ad2}}_{\ell} \\
    &\quad+ \alpha \hat{C}^{\mtr{iso}}_{\ell}
          + \mtr{sign}(\gamma)\sqrt{\alpha(1-\alpha)\abs{\gamma}}
            \hat{C}^{\mtr{cor}}_{\ell} \bigl] \\
    &\equiv C^{\mtr{ad1}}_{\ell} + C^{\mtr{ad2}}_{\ell} +
            C^{\mtr{iso}}_{\ell} + C^{\mtr{cor}}_{\ell},}
\end{equation}
where we have defined the total amplitude, the isocurvature fraction, and the
correlation at the pivot scale
\begin{eqnarray}
  A^{2} \equiv A^{2}_{r} + A^{2}_{s} + B^{2}, \qquad
  \alpha \equiv \frac{ B^{2} }{ A^{2} }
  \quad \alpha \in [0,1], \label{eq:alphadef}\\
  \gamma \equiv \mtr{sign}(A_{s}B)\frac{ A^{2}_{s} }{ A^{2}_{r} + A^{2}_{s} }
  \quad \gamma \in [-1,1]. \label{eq:gammadef}
\end{eqnarray}
The $\hat{C}_{\ell}$ denote spectra obtained with unit amplitude ($A_{r} = 1$,
$A_{s} = 1$, or $B = 1$).

Using this parametrization as a basis for a likelihood study has the problem,
that for small values of $\alpha$ or $\gamma$, the spectral indices $\niso$
and $\nadII$ become unconstrained. This slows down the convergence of the
analysis and, if flat priors for the spectral indices are used, may bias the
results toward zero $\alpha$ or $\gamma$.

Therefore, as suggested in \cite{Kurki-Suonio:2004mn}, we have switched to
using as primary parameters the amplitudes at two different scales $k_1 =
0.002~\mtr{Mpc}^{-1}$ and $k_2 = 0.05~\mtr{Mpc}^{-1}$, making the spectral
indices derived parameters. This parametrization and the related issue of
priors is discussed more in the Appendix.

In both parameterizations our model has 10 parameters. Four relate to the
background cosmology: the physical baryon density ($\omega_b = h^2\Omega_b$),
the physical CDM density ($\omega_c = h^2\Omega_c$), the sound horizon angle
($\theta$) and the optical depth to reionization ($\tau$). In the old
``index'' parametrization the six related to perturbations are the logarithm
of the overall primordial perturbation amplitude ($\ln A^2$) and the
fractional amplitudes $\alpha$ and $\gamma$, all taken at $k_0$, and the three
spectral indices $\nadI$, $\nadII$, and $\niso$.  In the new ``amplitude''
parametrization the six primary perturbation parameters are the amplitudes
$\ln A_1^2$, $\ln A_2^2$, $\alpha_1$, $\alpha_2$, $\gamma_1$, and $\gamma_2$,
taken at $k_1$ and $k_2$.

\section{Analysis}

We created 8 Monte Carlo Markov chains (MCMC) using the CosmoMC
\cite{Lewis:2002ah} engine, which we have modified to handle correlated
adiabatic and isocurvature modes. We accumulated a total of $1,130,306$ steps
with average multiplicity $10.1$, summing up to a total of $11,367,942$
samples.  The chains are well mixed, with the Gelman and Rubin
\cite{GelRub:1992} statistic $R-1 = 0.06$. In split ($1/2$) tests the 95\%
C.L. limits for the primary parameters change typically by less than 10\% of
the standard deviation, and even for the worst ones by less than 20\%.

Likelihood of a model was assessed using the WMAP 3-year data and likelihood
code (version v2p2p2) \cite{Spergel:2006hy, Page:2006hz, Hinshaw:2006ia},
Boomerang \cite{Jones:2005yb} and ACBAR \cite{Kuo:2002ua} data, and the LSS
data from the SDSS data release 4 luminous red galaxy sample (LRG)
\cite{Tegmark:2003uf, Tegmark:2006az}. To avoid the necessity of nonlinear
corrections to LSS growth, only the first $14$ $k$-bands were used. We
marginalize over the galaxy bias parameter and the ACBAR and Boomerang
calibration uncertainties.

Thus in our main analysis we use only four different data sets. The motivation
for keeping the number of different data sets small, is that with too many
data sets the interplay of their different systematic effects can become
problematic. We use the best available CMB and LSS data only. These are the
data where the acoustic peak structure can be observed and thus the data which
directly probe the adiabaticity of the perturbations. So the main object of
this study is to see whether the observed peak structure is that one which
characterizes adiabatic perturbations.

We assign flat (i.e., uniform over some range; see Appendix) prior
probabilities to our 10 primary parameters, using the amplitude
parametrization. Since some parameters are not tightly constrained by the
data, the posterior likelihoods depend somewhat on the prior probability
densities. To show the effect of the priors, and to compare to
\cite{Kurki-Suonio:2004mn}, we used the Jacobian of the parameter
transformation (see Appendix) to convert the 1-d marginalized likelihoods
obtained from our chains to correspond to flat priors for the index
parametrization (dotted lines in Fig.~\ref{fig:1dlikelihoods}).

\section{Results}

\begin{figure*}
  \centering
  \includegraphics*[width=\textwidth]{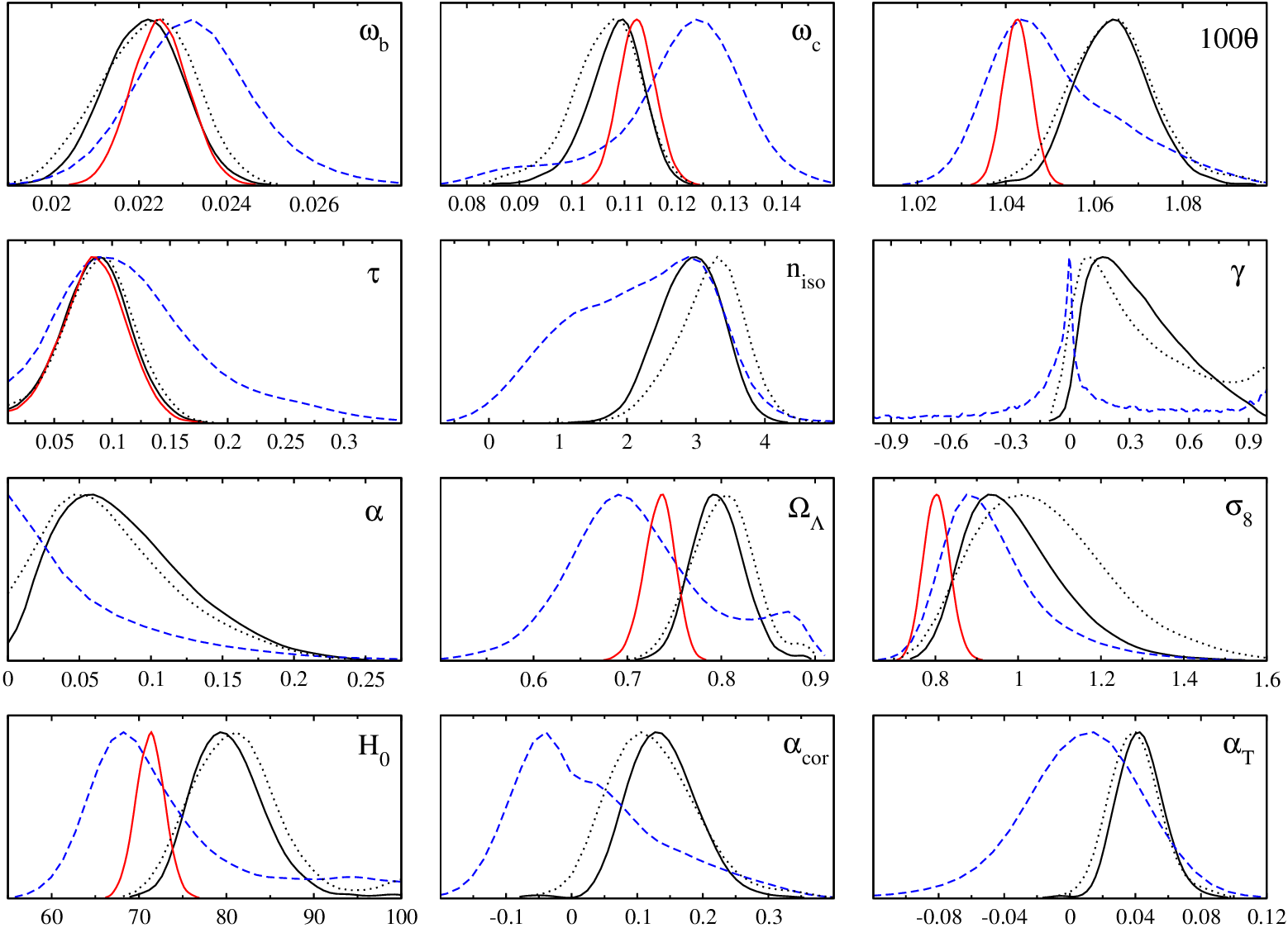}
  \caption{Marginalized likelihood functions for selected primary and derived
    parameters. The {\em solid black} curves are our new results using WMAP
    3-year data and Boomerang data from the 2003 flight (+ ACBAR \& SDSS).
    The {\em dotted black} curves show the effect of assigning flat priors in
    the index parametrization instead of the amplitude parametrization.  The
    {\em red/gray} curves are for an adiabatic model using the same data.  The
    {\em dashed blue} curves are from our previous study
    \protect\cite{Kurki-Suonio:2004mn} using data available in 2004.  Note
    that also in the adiabatic model WMAP3 data favors a larger $H_0$ than
    WMAP1 (not shown)---allowing isocurvature modes favors larger $H_0$
    regardless of which WMAP data set is used, although the effect is much
    stronger with WMAP3.}
  \label{fig:1dlikelihoods}
\end{figure*}

\begin{table*}
  \caption{\label{tab1}The best-fit models. A: The full model with a
    correlated isocurvature mode. B: The adiabatic model.}
  \begin{indented}
  \item[]\begin{tabular}{llllllllll}
      \br
      & $\omega_b$ & $\omega_c$ & $100\theta$ & $\tau$ & $\nadI$ & $\nadII$
      & $\niso$ & $\alpha$ & $\gamma$ \\
      \mr
      A & 0.0223 & 0.1066 & 1.062 & 0.0914 & 0.975 & 0.919 & 3.54 & 0.0539 & 0.180 \\
      B & 0.0224 & 0.1124 & 1.042 & 0.0856 & 0.960 \\
      \br
    \end{tabular}
  \item[]\begin{tabular}{llllll}
      \br
      & $\Omega_\Lambda$ & $H_0$ & $\sigma_8$ & $\acor$ & $\alpha_T$ \\
      \mr
      A & 0.800 & 80.3 & 1.11 & 0.096 & 0.0356 \\
      B & 0.734 & 71.2 & 0.80 \\
      \br
    \end{tabular}
  \end{indented}
\end{table*}

We show 1-d marginalized likelihoods for selected primary and derived
parameters in Fig.~\ref{fig:1dlikelihoods}.  Instead of the amplitudes at
$k_1$ and $k_2$ we show the derived parameters: spectral index $\niso$ and the
amplitude parameters $\alpha$ and $\gamma$ at the intermediate scale $k_0 =
0.01 \mbox{Mpc}^{-1}$. The mappings between these are presented in the
Appendix.

The data lead to likelihood peaks at clearly nonzero values for $\alpha$ (the
ratio of the primordial entropy perturbation power to the total perturbation
power at $k_0$) and favor a positive $\gamma$. (This significantly reduces the
pivot-scale dependence of the likelihoods discussed in
\cite{Kurki-Suonio:2004mn}.) If the subdominant entropy perturbation is
correlated with the dominant curvature perturbation it has a much stronger
effect on the observables. To account for this we also show $\acor \equiv
\mtr{sign}(\gamma)\sqrt{\alpha(1-\alpha)\abs{\gamma}}$, which gives the
relative ``weight'' of the correlation spectrum $C_\ell^\mtr{cor}$ in
$C_\ell$, see \Eref{eq:total-Cl}. We find $\alpha = 0.082\pm0.046$ (mean $\pm$
stdev), with $\alpha < 0.169$ at 95\% C.L.\ and $\acor = 0.139\pm0.055$,
with $0.044 < \acor < 0.252$ at 95\% C.L. The correlation between
the isocurvature and adiabatic modes is positive, $\gamma > 0.051$
at 95\% C.L.

These values are obtained from the likelihood functions corresponding to the
amplitude parametrization with flat priors for the primary parameters.  As
can be seen from Fig.~\ref{fig:1dlikelihoods} the exact confidence levels
depend on the assumed priors, but the conclusions are not changed at a
qualitative level. We discuss the priors more in the Appendix.

Since the definitions of the primary amplitude parameters (e.g. $\alpha$)
depend on the choice of pivot scale, we also define
 \begin{equation}
   \alpha_T \equiv
   \frac{\sum(2\ell+1)(C_\ell^\mtr{iso}+C_\ell^{\mtr{cor}})}
   {\sum(2\ell+1)C_\ell} \,,
 \end{equation}
the total nonadiabatic contribution to the CMB temperature variance,
\begin{equation}
  \Bigg\langle\left(\frac{\delta T}{T}\right)^2\Bigg\rangle
  = \sum_\ell \frac{2\ell+1}{4\pi}C_\ell \,.
\end{equation}
We find $\alpha_T = 0.043\pm0.015$, and the whole 95\% C.L. range $0.017 <
\alpha_T < 0.073$ is positive. Thus the CMB data clearly favor a nonadiabatic
contribution.

\begin{figure}
  \centering
  \includegraphics*[width=\textwidth]{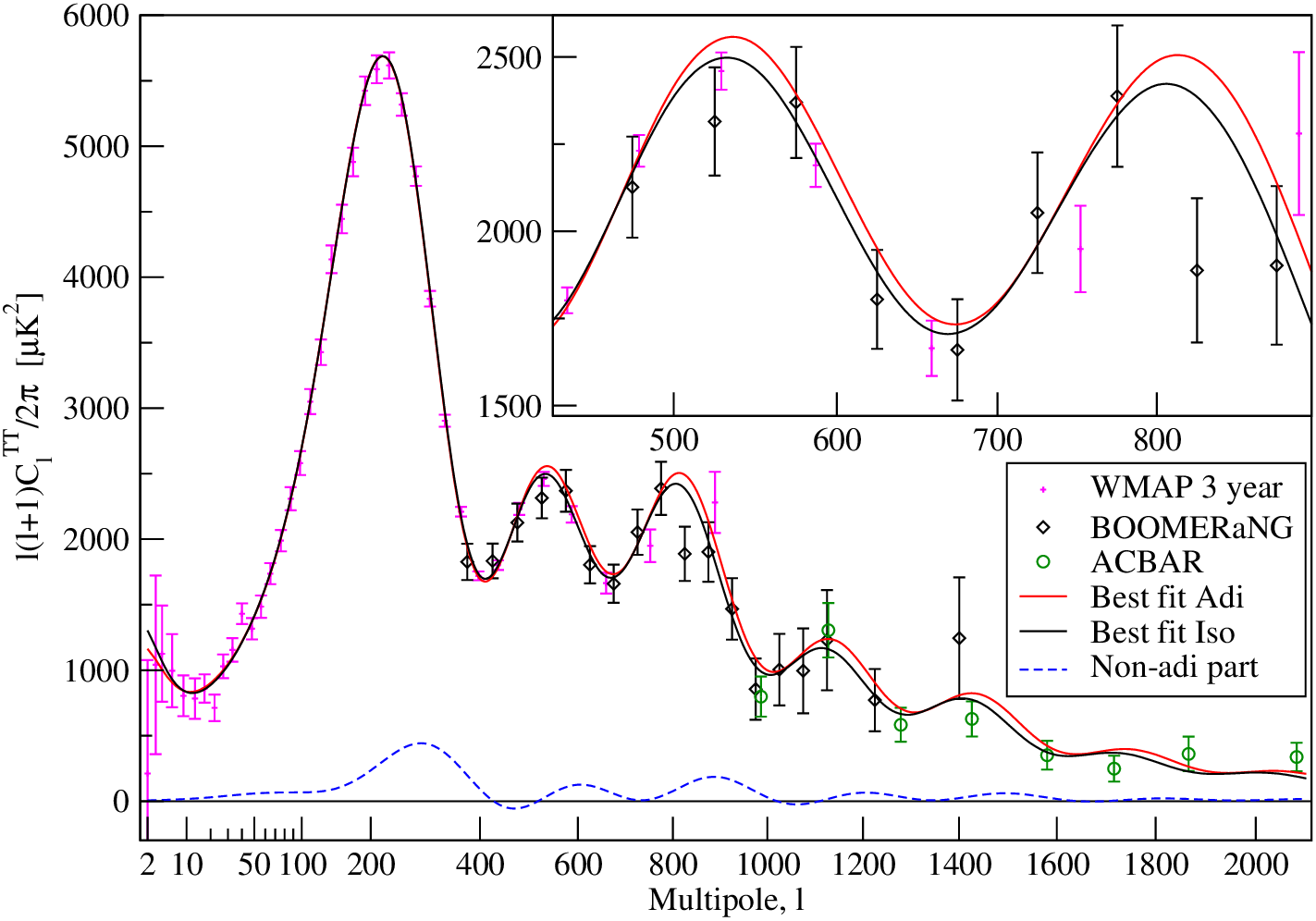}
  \caption{The CMB temperature angular power spectrum for our best-fit model
    (\emph{black}) compared to the best-fit adiabatic model (\emph{red/gray}).
    The \emph{dashed blue} curve shows the nonadiabatic contribution. The
    inset shows the 2nd and 3rd peaks.}
  \label{fig:bestfitCl}
\end{figure}

Our best-fit model and also the best-fit adiabatic model are given in Table I.
They are compared in Fig.~\ref{fig:bestfitCl}. The WMAP 3-year data prefer a
slightly narrower 2nd peak than the adiabatic model can produce. This holds
for the third peak also, but now the peak position and width are determined by
the Boomerang data.

This feature in the data can be accounted for by a correlated isocurvature
component. It can narrow down the 2nd peak without affecting the 1st peak
position, which is accurately determined by the data. Increasing $\theta$
shifts the whole peak structure to the left, making the peaks narrower. Adding
a positively correlated isocurvature component returns the 1st peak to its
place.

Compared to the adiabatic model, adding 4 parameters improved the fit by
$\Delta\chi^{2} = 9.7$. This comes mainly from the fit to the WMAP and
Boomerang temperature (TT) $C_\ell$; for the WMAP part from the 2nd peak. The
data on the TE and EE $C_\ell$ are too inaccurate for a few per cent
isocurvature contribution to play any role. Therefore the contribution to
$\chi^2$ from these cross-correlation data is the same for the adabatic and
our model. The fit to the SDSS data has improved slightly, whereas that to
the ACBAR data is indifferent. In Table~\ref{tab2} we give quantitative numbers
for the contributions of different data sets to the best-fit models. The
best-fit model and its $\chi^2$ with the data are of course not affected by the
issue of priors, which affects only the likelihoods. For the first time in the
history of CMB temperature anisotropy observations, adding the CDM (or baryon)
isocurvature degree of freedom improves the fit to the data, and the likelihoods
of isocurvature parameters peak at non-zero values, as seen in Tables~\ref{tab1}
\& \ref{tab2} and Fig.~\ref{fig:1dlikelihoods}.

\begin{table*}
  \caption{\label{tab2}The $\chi^2$ of the fit of the best-fit models to the data,
    and the contributions from the four different data sets. A: The full
    model with a correlated isocurvature mode. B: The adiabatic model.}
  \begin{indented}
  \item[]\begin{tabular}{lrrr}
      \br
      data set & $\chi^2$ (A) & $\chi^2$ (B) & $\Delta \chi^2$  \\
      \mr
      WMAP & 3535.20 & 3539.14 & 3.94  \\
      Boomerang & 31.12 & 35.04 & 3.92  \\
      ACBAR & 10.10 & 9.92 & $-0.18$ \\
      SDSS LRG & 22.64 & 24.66 & 2.02 \\
      \mr
      total & 3599.06 & 3608.78 & 9.72 \\
      \br
    \end{tabular}
  \end{indented}
\end{table*}

The other major effect is that the correlated isocurvature model favors a
smaller CDM density $\omega_c$ (also $\omega_b$ is down, but not as much).
This is due to the correlation component $C_\ell^\mtr{cor}$, which raises the
first and third peaks with respect to the second one. See
Fig.~\ref{fig:bestfitCl}.  Lower $\omega_m$ and $\omega_b$ compensate this by
raising the second peak.

For fixed $\omega_c$ and $\omega_b$, increasing $\theta$ leads to a larger
$\Omega_\Lambda$ and a larger Hubble constant $H_0$.  For fixed $\theta$ and
$\omega_b$, a lower $\omega_c$ requires an even larger $\Omega_\Lambda$ and
$H_0$.  Therefore these models have a larger $H_0$ and a smaller matter
density parameter $\Omega_m = 1 - \Omega_\Lambda$, than the adiabatic model.

\section{Other cosmological data}

\begin{figure*}
  \centering
  \includegraphics*[width=\textwidth]{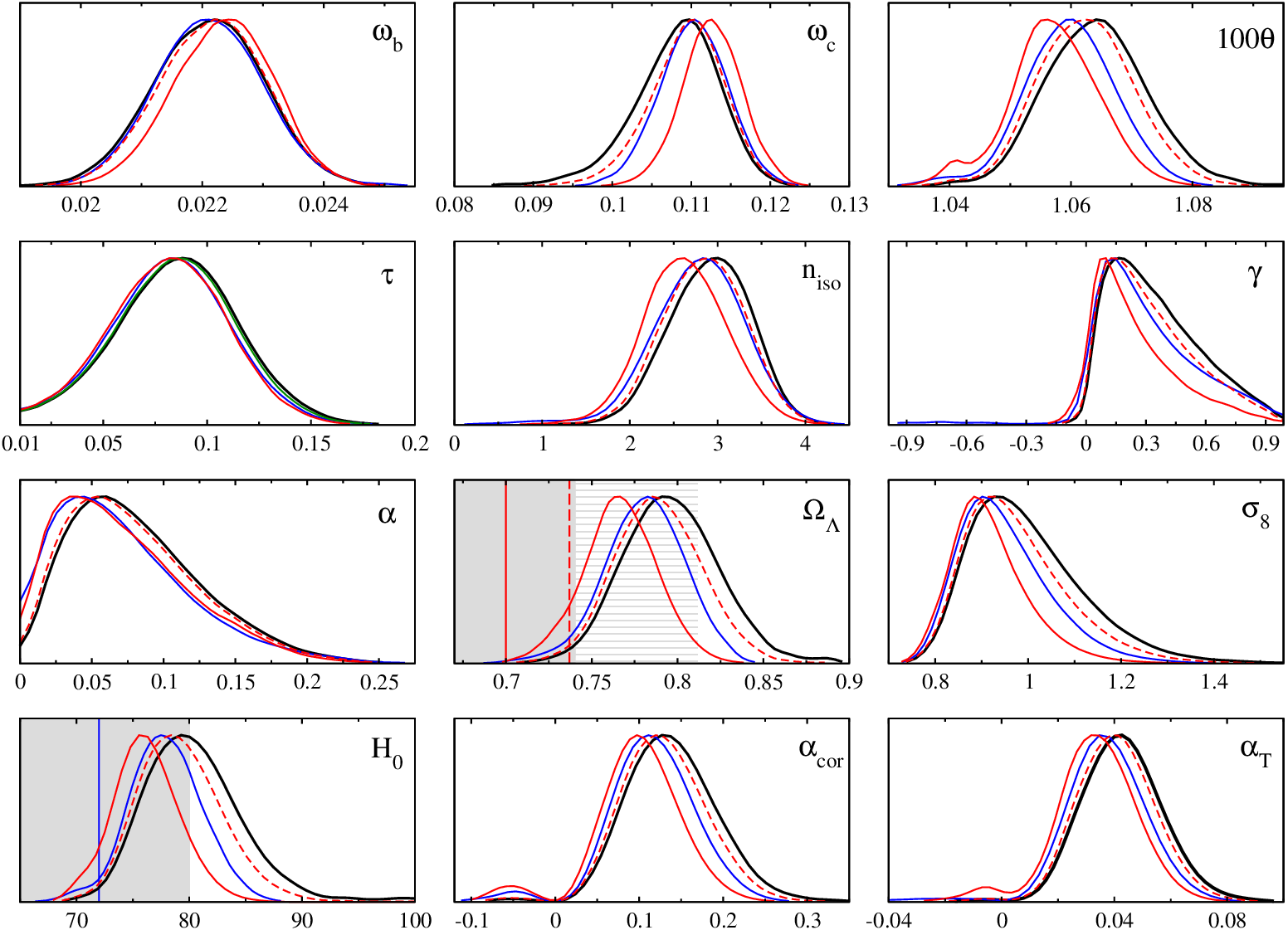}
  \caption{Marginalized likelihood functions for selected primary and derived
    parameters after including additional priors coming from other
    cosmological data.  The {\em solid black} curves are the same as in
    Fig.\ref{fig:1dlikelihoods}.  The {\em solid blue} curves show the effect
    of adding the HST $H_0$ prior. The {\em solid red} curves show the effect
    of the stronger $\Omega_m = 0.29^{+0.05}_{-0.03}$ prior and the {\em
      dashed red} curves are for the weaker $\Omega_m = 0.263\pm0.074$ prior.
    The central values and 1-$\sigma$ ranges of these Gaussian priors are
    indicated by vertical lines and {\em gray} areas in the $\Omega_\Lambda$
    and $H_0$ panels.}
  \label{fig:othercosm}
\end{figure*}

The large $H_0 = 80\pm4\mtr{ km/s/Mpc}$ and small $\Omega_m = 0.204\pm0.028$
of our nonadiabatic model is in some tension with other cosmological data,
like the Hubble Space Telescope (HST) value for the Hubble constant, $H_0 =
72\pm 8 \mtr{ km/s/Mpc}$ \cite{Freedman:2000cf} and the estimates of
$\Omega_m$ from Supernova Ia data ($\Omega_m = 0.29^{+0.05}_{-0.03}$ from
\cite{Riess:2004nr}, or $\Omega_m = 0.263\pm0.042\pm0.032$ from
\cite{Astier:2005qq}). To assess the effect of an $\Omega_m$ constraint, we
postprocessed our likelihoods by weighting (importance sampling
\cite{Lewis:2002ah}) our MCMC chains with corresponding Gaussian distributions
for $\Omega_m$.  We also ran new MCMC chains with a Gaussian $H_0 = 72\pm 8
\mtr{ km/s/Mpc}$ \cite{Freedman:2000cf} prior. The results are shown in
Fig.~\ref{fig:othercosm}.  The effect of the weaker $\Omega_m = 0.263\pm0.074$
prior is minor and does not eliminate the favoring of nonzero $\alpha$ and
positive $\alpha_T$. The same is true for the $H_0$ prior, although now the
95\% C.L. range $-0.005 < \alpha_T < 0.065$ includes $\alpha_T = 0$. An
$\Omega_m = 0.30\pm0.04$ prior has a larger effect, but the 1-d likelihood of
$\alpha$ appears still to peak at a nonzero value.

As noted in \cite{Kurki-Suonio:2004mn}, $\niso \simeq 3$ (when $\nad \simeq
1$) means that the relative isocurvature contribution to $C_\ell$ is about the
same on all scales, but in the LSS power spectrum the isocurvature and
correlation components overtake the adiabatic ones at small scales. Thus there
is more (matter) power at small scales, with a larger rms mass fluctuation on
8 $h^{-1}$Mpc scale $\sigma_8$. While the adiabatic model favors $\sigma_8
\approx 0.8$, our model favors $\sigma_8 \approx 1$. Assuming the power law is
a good approximation for the primordial spectra also at smaller scales,
Lyman-$\alpha$ data can constrain $\niso$ \cite{Beltran:2005gr}. Creating new
MCMC chains with Lyman-$\alpha$ data included like in \cite{Beltran:2005gr} we
checked that the peak in the likelihood of $\niso$ shifted to $2.1$, and
$\alpha$ to zero.  The other likelihoods became more like in the adiabatic
case. The peak in $\alpha_T$ remained clearly positive, but the 68\% C.L.
region, $\alpha_T = 0.022^{+0.017}_{-0.025}$, now included $\alpha_T = 0$.
Thus the large small-scale power implied by a relatively large constant
$\niso$ means that inclusion of small-scale data significantly weakens the
preference for a nonadiabatic contribution. One possible interpretation is
that $\niso$ is not a constant, but becomes smaller for the large $k$ probed
by Lyman-$\alpha$ data.

\section{Discussion}

The acoustic peak structure in the latest CMB data favors a few per cent
nonadiabatic contribution. Our best fit to the CMB + LSS data has a
significant contribution from a positively correlated isocurvature mode, so
that the isocurvature fraction, i.e., the ratio of primordial entropy
perturbation power to total perturbation power at the scale of $k = 0.01
\mtr{~Mpc}^{-1}$, is 5\% and the nonadiabatic contribution to CMB temperature
variance is 4\%.  With the addition of four model parameters, the fit to the
data improved by $\Delta\chi^2 = 9.7$, which can already be considered
interesting.

We note that the nonadiabatic models that give the best fits to the CMB data,
have some problems with other cosmological data. These problems may be
remedied if one considers non-flat models. A preliminary look at these models
seems to indicate that the best-fit non-flat models are closed models with a
total energy density a few per cent over the critical density, a Hubble
constant close to the HST value, $\Omega_m$ and $\Omega_\Lambda$ in agreement
with the SNIa data, the isocurvature spectral index $\niso$ closer to 2 than
3, and an even larger nonadiabatic contribution than in the case of flat
nonadiabatic models.

While interesting, the preference for nonadiabatic primordial perturbations
discovered in this study should not be taken as conclusive evidence for a
primordial isocurvature mode.  It is not yet at a very high confidence level,
so the possibility remains that it is just a statistical fluke. Other possible
explanations are an unaccounted for systematic effect in the WMAP and
Boomerang data, or some other nonstandard cosmological feature whose effect on
the CMB is in a similar direction. It remains for the future more precise CMB
measurements to resolve this.  If the presence of a primordial isocurvature
mode were confirmed, it would automatically rule out \emph{single-field}
inflation, since generating a primordial entropy perturbation requires more
than one degree of freedom.

\ack We thank the CSC - Scientific Computing Ltd.  (Finland) for computational
resources. JV thanks C.\ Byrnes for useful discussions. We are grateful to M.\
Beltr\'{a}n, J.\ Garc\'{i}a-Bellido \& J.\ Lesgourgues for their comments. We
acknowledge the use of the Legacy Archive for Microwave Background Data
Analysis (LAMBDA) (NASA). RK is supported by the Jenny and Antti Wihuri
Foundation. VM is supported by the Finnish Graduate School in Astronomy and
Space Physics. VM thanks the Galileo Galilei Institute for Theoretical Physics
for hospitality and INFN for partial support. JV is supported by PPARC grant
PP/C502514/1 and the Academy of Finland grants 112383 \& 120181. This work was
supported by the Academy of Finland grant No. 205800 and by the European Union
through the Marie Curie Research and Training Network ``UniverseNet''
(MRTN-CT-2006-035863).

\appendix
\section{Priors and parametrizations}

When some parameters of a model are not sufficiently tightly constrained by
the data, the posterior likelihood functions become sensitive to the assumed
prior probability densities for the parameters. Even when one assumes flat,
i.e., uniform, priors for the primary parameters of the model, the question
remains, which parameters are taken to be the primary parameters, since the
priors for the quantities derived from the primary parameters (derived
parameters) will not be flat.  To avoid problems related to spectral indices
becoming unconstrained when the corresponding amplitude parameters have small
values, we introduced in this study a parametrization in terms of amplitudes
at two different scales, instead of using spectral indices as primary
parameters.

Mapping from the amplitude parametrization to the spectral index
parametrization is easy to find by employing definitions
(\ref{eq:correlations}), (\ref{eq:alphadef}), and (\ref{eq:gammadef}). The
spectral indices can be written in terms of the parameters of amplitude
parametrization as
\begin{eqnarray}
  \nadI - 1 = \frac{ \ln\left[\mtc{P}_{\mtc{R}1}(k_2) /
      \mtc{P}_{\mtc{R}1}(k_1)\right] }
  {\ln(k_2/k_1)}\,,\label{eq:newnadI} \\
  \nadII - 1 = \frac{ \ln\left[\mtc{P}_{\mtc{R}2}(k_2) /
      \mtc{P}_{\mtc{R}2}(k_1)\right] }
  {\ln(k_2/k_1)}\,,\label{eq:newnadII} \\
  \niso - 1 = \frac{ \ln\left[\mtc{P}_{\mtc{S}}(k_2) /
      \mtc{P}_{\mtc{S}}(k_1)\right] }
  {\ln(k_2/k_1)}\,,\label{eq:newniso}
\end{eqnarray}
where the first (uncorrelated) adiabatic, the second (correlated) adiabatic
and the isocurvature power spectra at scales $k_i$ ($i=1,2$) are given by
\begin{eqnarray}
  \mtc{P}_{\mtc{R}1}(k_i) = A_i^2 (1 - \alpha_i) (1-\abs{\gamma_i})\,,\\
  \mtc{P}_{\mtc{R}2}(k_i) = A_i^2 (1 - \alpha_i) \abs{\gamma_i}\,,\\
  \mtc{P}_{\mtc{S}}(k_i) = A_i^2 \alpha_i\,,\label{eq:PS}
\end{eqnarray}
respectively. Then the amplitudes $A$, $\alpha$, and $\gamma$ at $k_0 =
0.01\;$Mpc$^{-1}$ are obtained from the amplitude-parametrization amplitudes
$A_1$, $\alpha_1$, and $\gamma_1$ defined at $k_1 = 0.002\;$Mpc$^{-1}$ by
\cite{Kurki-Suonio:2004mn}
\begin{eqnarray}
  A^{2} = A_1^{2}\Big[(1-\alpha_1)(1-\abs{\gamma_1})\tilde{k}^{\nadI-1} +
  (1-\alpha_1)\abs{\gamma_1}\tilde{k}^{\nadII-1} +
  \alpha_1\tilde{k}^{\niso-1}\Big] \,,\label{eq:newA} \\
  \alpha = \frac{\alpha_1\tilde{k}^{\niso-1}}
  {(1-\alpha_1)(1-\abs{\gamma_1})\tilde{k}^{\nadI-1} +
    (1-\alpha_1)\abs{\gamma_1}\tilde{k}^{\nadII-1} +
    \alpha_1\tilde{k}^{\niso-1}}\,, \label{eq:newalpha} \\
  \gamma = \frac{\gamma_1\tilde{k}^{\nadII-1}}
  {(1-\abs{\gamma_1})\tilde{k}^{\nadI-1} +
    \abs{\gamma_1}\tilde{k}^{\nadII-1}}\,, \label{eq:newgamma}
\end{eqnarray}
where $\tilde{k} = k_0/k_1$, and the spectral indices are obtained from
Eqs.~(\ref{eq:newnadI})--(\ref{eq:newniso}).

If we take as a starting point the spectral index parametrization, then the
mapping from it to the amplitude parametrization is simply
\begin{eqnarray}
  A_i^{2} = A^{2}\Big[(1-\alpha)(1-\abs{\gamma})\tilde{k}^{\nadI-1} +
  (1-\alpha)\abs{\gamma}\tilde{k}^{\nadII-1} +
  \alpha\tilde{k}^{\niso-1}\Big] \,,\label{eq:newA2}\\
  \alpha_i = \frac{\alpha\tilde{k}^{\niso-1}}
  {(1-\alpha)(1-\abs{\gamma})\tilde{k}^{\nadI-1} +
    (1-\alpha)\abs{\gamma}\tilde{k}^{\nadII-1} +
    \alpha\tilde{k}^{\niso-1}}\,, \label{eq:newalpha2}\\
  \gamma_i = \frac{\gamma\tilde{k}^{\nadII-1}}
  {(1-\abs{\gamma})\tilde{k}^{\nadI-1} +
    \abs{\gamma}\tilde{k}^{\nadII-1}}\,, \label{eq:newgamma2}
\end{eqnarray}
where $\tilde{k} = k_i/k_0$ and $i=1,2$.

Since we assume that all the component spectra can be described by power laws,
$\gamma_1$ and $\gamma_2$ must have the same sign.  Hence, they are not
completely independent. To obtain independent primary parameters, we draw
$\gamma_1$ from the range $[-1,1]$, but $\gamma_2$ only from the range
$[0,1]$, and let $\gamma_1$ determine the sign of the correlation. This has no
effect on formulas (\ref{eq:newnadI}) -- (\ref{eq:newalpha2}), but to maintain
consistency we need to modify equation (\ref{eq:newgamma2}). Instead of
$\gamma$ we should have in the nominator $\abs{\gamma}$, if $i=2$.

Employing mappings (\ref{eq:newnadI}) -- (\ref{eq:newniso}) and
(\ref{eq:newA}) -- (\ref{eq:newgamma}) we obtain the posterior likelihoods of
$\nadI$, $\nadII$, $\niso$, $A$, $\alpha$, $\gamma$ for a MCMC run in the
amplitude parametrization (corresponding to flat priors for $A_1$, $\alpha_1$,
$\gamma_1$, $A_2$, $\alpha_2$, and $\gamma_2$). Likewise, if a MCMC run was
made in spectral index parametrization, we could use mappings (\ref{eq:newA2})
-- (\ref{eq:newgamma2}) to obtain the posterior likelihoods of $A_1$,
$\alpha_1$, $\gamma_1$, $A_2$, $\alpha_2$, $\gamma_2$ (and other parameters)
corresponding to flat prior probabilities for the primary parameters of the
index parametrization.

However, if we want to convert the results obtained in the amplitude
parametrization to flat priors for the spectral indices, then the mapping
(\ref{eq:newnadI}) -- (\ref{eq:newniso}) and (\ref{eq:newA}) --
(\ref{eq:newgamma}) is not enough. We have to correct for the prior too. This
can be done by weighting the multiplicities in the MCMC chains (i.e. weighting
the posterior likelihood) by the Jacobian of the transformation
(\ref{eq:newnadI}) -- (\ref{eq:newniso}) and (\ref{eq:newA}) --
(\ref{eq:newgamma}). If the MCMC run was made using primary parameters $\{
\Theta_i \}$ (and flat priors for them), but we want to show the results with
flat priors for $\{ \tilde\Theta_i \}$, the multiplicities must be multiplied
by
\begin{equation}
  J = \left|\det\left(\frac{\partial\Theta_i}{\partial\tilde\Theta_j}\right)\right|\,.
  \label{eqn:jacobian}
\end{equation}

\begin{figure*}
  \centering
  \includegraphics*[width=\textwidth]{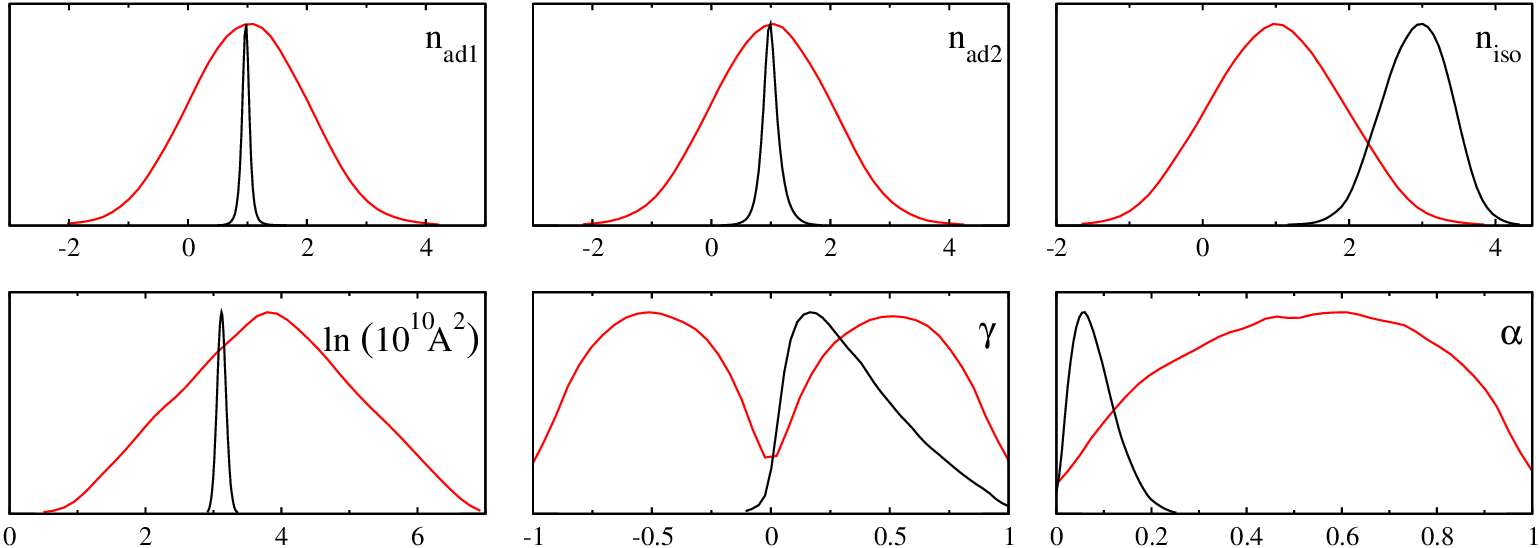}
  \caption{The {\em red} curves show the priors for the parameters of the
    index parametrization, when flat priors are assumed for the parameters of
    the amplitude parametrization.  The {\em black} curves are the posterior
    likelihoods we have obtained (the same as the solid black curves in
    Fig.~\ref{fig:1dlikelihoods}).  }
  \label{fig:aprior}
\end{figure*}

\begin{figure*}
  \centering
  \includegraphics*[width=\textwidth]{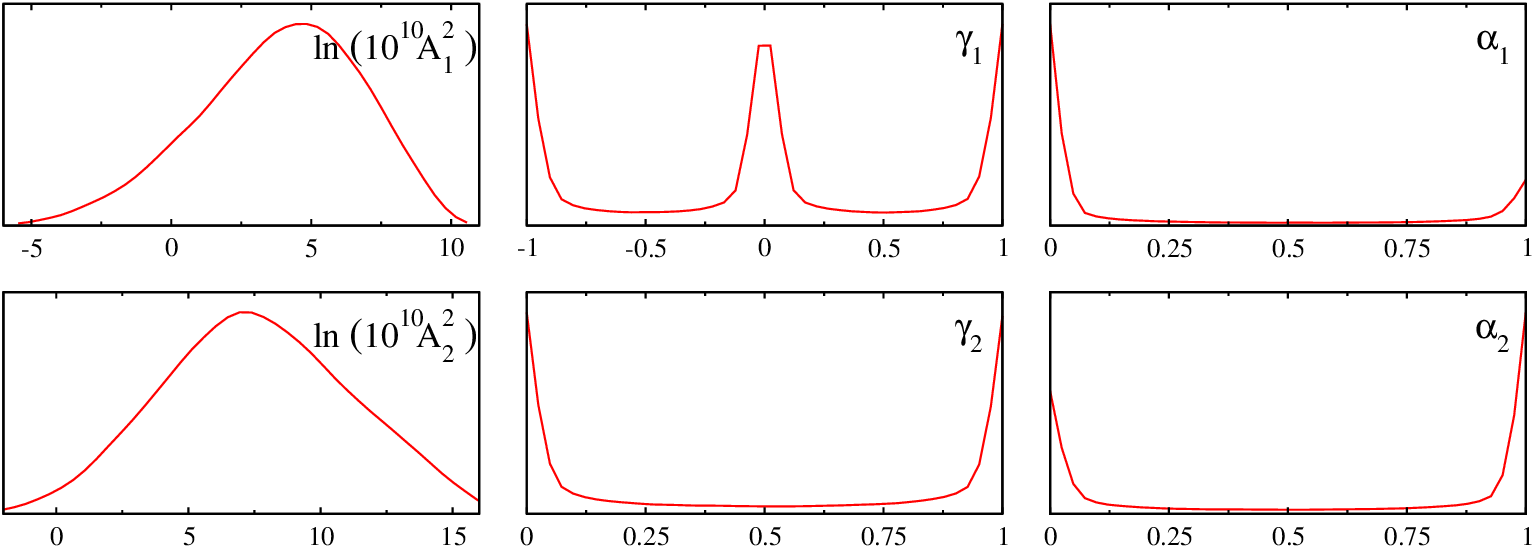}
  \caption{The priors for the parameters of the amplitude parametrization,
    when flat priors are assumed for the parameters of the index
    parametrization.  }
  \label{fig:nprior}
\end{figure*}

In Figs.~\ref{fig:aprior} and \ref{fig:nprior} we show what the priors of one
parametrization become when the priors of the other parametrization are
taken to be flat.

In the amplitude parametrization the range of parameters (which is also a
crucial part of the prior) is throughout our study
\begin{eqnarray}
  \omega_b \in [0.005, 0.1],\quad \omega_c \in [0.01, 0.99],
  \quad 100\theta \in [0.3, 10.0], \quad  \tau \in [0.01, 0.3] \nonumber \\
  \ln(10^{10}A_1^{2}) \in [1, 7], \quad
  \alpha_1 \in [0, 1],\quad \gamma_1 \in [-1, 1]\nonumber\\
  \ln(10^{10}A_2^{2}) \in [1, 7], \quad \alpha_2 \in [0, 1], \quad
  \gamma_2 \in [0, 1]\,.\nonumber
\end{eqnarray}
In the spectral index parametrization the range of background parameters is
the same as above, and the perturbation parameters have a uniform prior
probability over the ranges
\begin{eqnarray}
  \nadI \in [-3, 4],\quad \nadII \in [-3, 4], \quad \niso \in [-3, 12],\nonumber\\
  \ln(10^{10}A^{2}) \in [1, 7], \quad \alpha \in [0, 1], \quad
 \gamma \in [-1, 1]\,. \nonumber
\end{eqnarray}
These are used only for producing Fig.~\ref{fig:nprior}, since all our
analysis presented in this paper is based on MCMC runs in the amplitude
parametrization. However, we have checked against a MCMC run using the index
parametrization, that its results agree with the dotted black curves in
Fig.~\ref{fig:1dlikelihoods} (which were obtained from the MCMC run using
amplitude parametrization by correcting for the prior with the Jacobian
(\ref{eqn:jacobian})).  For Fig.~\ref{fig:nprior} we chose the above ranges of
spectral indices to match this check run and our previous study
\cite{Kurki-Suonio:2004mn}. As indices are not symmetric around the scale
invariance ($n=1$), we can see mild asymmetries in Fig.~\ref{fig:nprior}.

\begin{figure*}
  \centering
  \includegraphics*[width=0.8\textwidth]{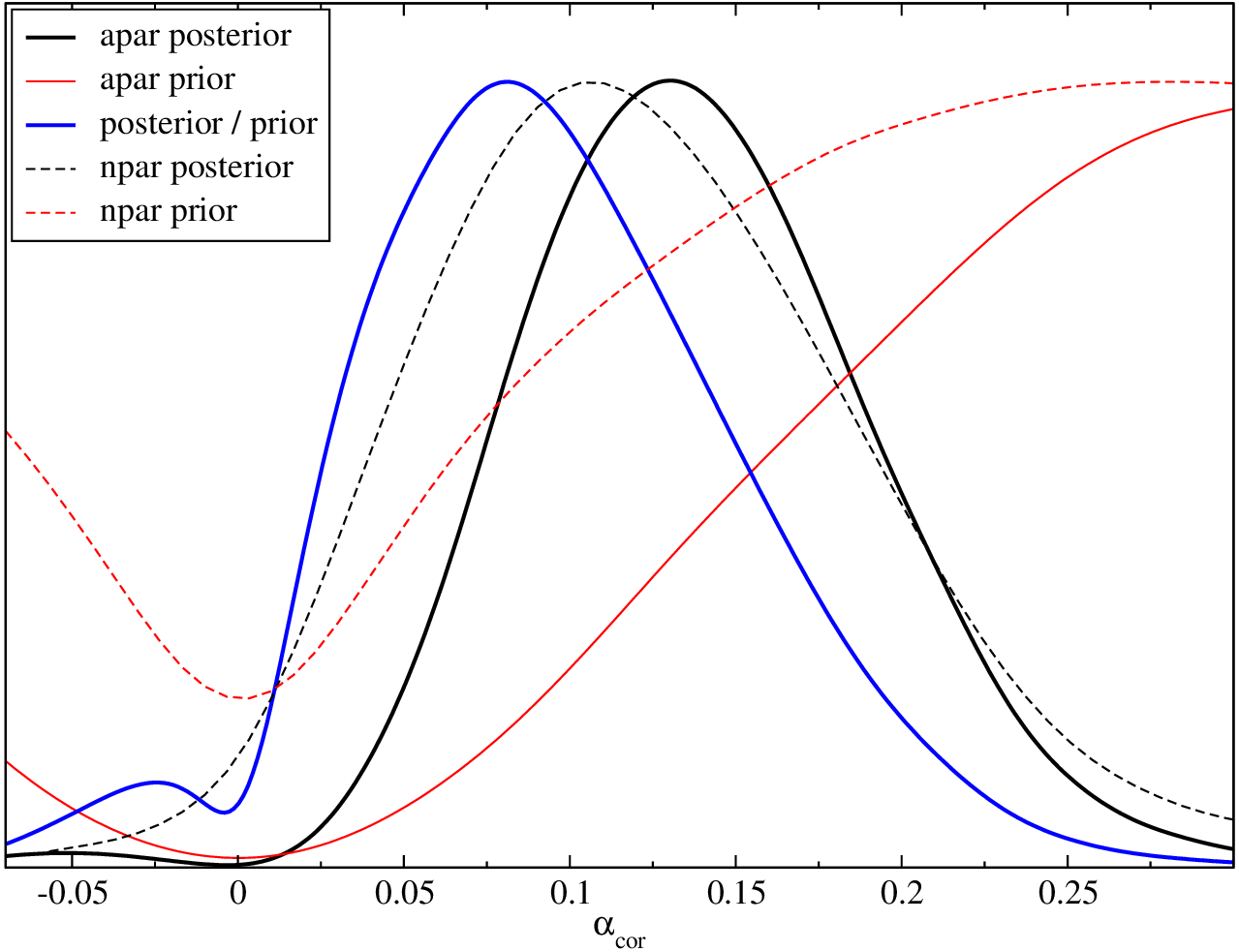}
  \caption{Prior and posterior likelihoods for the derived parameter $\acor$.
    The {\em solid red} curve is the prior in the amplitude parametrization
    and the {\em dashed red} curve is the same in the index parametrization.
    The {\em solid} and {\em dashed black} curves are the corresponding
    posterior likelihoods. These are the same as the solid and dotted black
    curves in the $\acor$ panel of Fig.~\ref{fig:1dlikelihoods}. The {\em
      solid blue} curve is the ratio of the posterior and prior likelihoods in
    the amplitude parametrization. }
  \label{fig:acor}
\end{figure*}

The parameters $\acor$ and $\alpha_T$ are derived parameters and therefore their
prior distributions are not flat in either parametrization.  We demonstrate the
situation with $\acor$ in Fig.~\ref{fig:acor}. We see that $\acor$ as well as
the parameters $\niso$, $\gamma$, and $\alpha$ (see Fig.~\ref{fig:aprior}) are
not well enough determined by the data to make their likelihood functions
insensitive to the choice of priors. The solid blue curve in Fig.~\ref{fig:acor}
is the ratio of the posterior and prior likelihoods for $\acor$, and illustrates
what the likelihood of $\acor$ could be, if $\acor$ had a flat prior probability
density. The actual likelihood of course depends also on the other priors.

\begin{figure*}
  \centering
  \includegraphics*[width=0.8\textwidth]{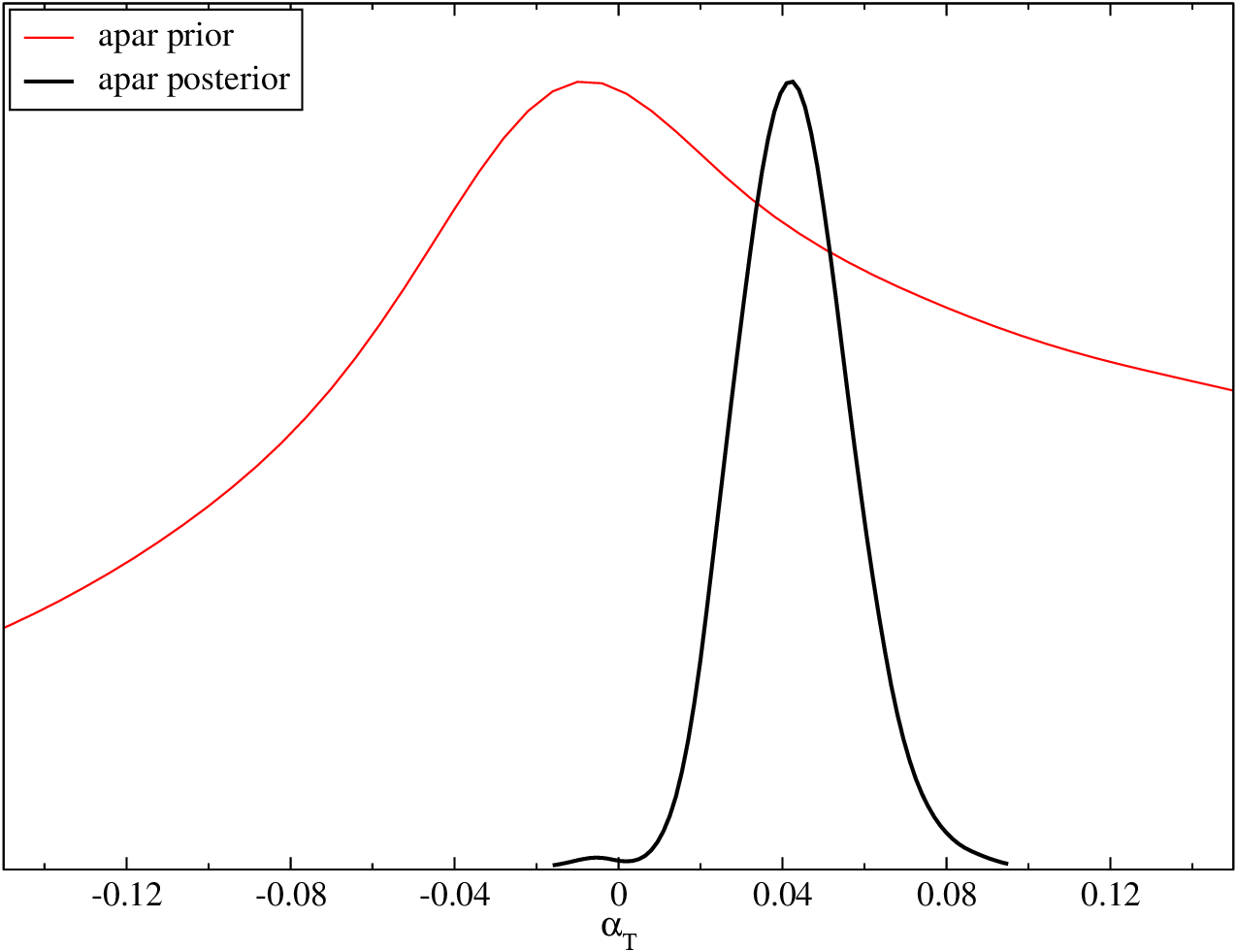}
  \caption{Prior {\em thin red} and posterior {\em thick black} likelihoods for
  the derived parameter $\alpha_T$. Both likelihoods are for the amplitude
  parametrization. Posterior/prior would be almost indistinguishable from the
  posterior.}
  \label{fig:alphaT}
\end{figure*}

In Fig.~\ref{fig:alphaT} we show the prior and posterior distributions of
$\alpha_T$ in the amplitude parametrization.  We see that the prior is
relatively flat in the region where the posterior mostly lies. Consequently,
unlike for $\acor$ the ratio of posterior and prior likelihoods
(posterior/prior) for $\alpha_T$ is almost indistinguishable from the posterior.
Thus the situation with $\alpha_T$ is better in this respect than the case of
$\acor$, and the conclusions about $\alpha_T$ are more robust.

\pagebreak
\section*{References}
\bibliographystyle{JHEP}
\bibliography{IsoH_JCAPv3}

\end{document}